\documentstyle[11pt]{article}

\newcommand{\p}{$\;\;$}

\begin{document}

\begin{center}
\vspace*{15mm}
{\Large\bf BINARY GLOBULAR CLUSTERS} \\[10mm]
{\large\bf G.A.Gurzadyan }\\[15mm]
{\bf Garny Space Astronomy Institute,  PO Box 370/15, Yerevan 2, Armenia}\\[20mm]
\end{center}

{\small\noindent
{\bf Abstract.} The dynamics of a close binary system of globular
clusters is considered. It is shown that the star transfer process
from one of the components to the other should lead to the decrease
of dimension of the first cluster with simultaneous increase of the
dimension of the second cluster. The rates of variations of clusters'
sizes due to the mutual tidal interaction between the components, as
well as of the orbital rotation of the binary system on the evolution
of clusters' sizes, are estimated. The results may be attributed to
the double nuclei observed in certain galaxies.
}

\section{The statement of the Problem}       %1

    The contemporary views on the evolution of globular clusters include
the following effects:\\[-7mm]
\begin{itemize}
\item [ i.]  Evaporation of stars. \\[-7mm]
\item [ ii.] The radius (tidal) of a cluster is determined by the condition
of compensation of the cluster's gravity and the field of the Galaxy.\\[-7mm]
\item [ iii.] Core collapse.\\[-7mm]
\item [ iv.] Core collapse may be halted by
one or several hard binaries.\\[-7mm]
\item [ v.] Tidal interaction or shocks of the Galaxy can accelerate the
the core collapse.\\[-7mm]
\end{itemize}

      The time scales of these processes for typical globular clusters are
of the order of billion years.

      Consider now a binary system both components of which are globular
clusters, i.e. when we have a {\it binary globular cluster}, particulary a
{\it cose binary globular cluster}.
Our problem here is to reveal heuristically the principal effects  which
arise when two ordinary globular clusters can be components of a binary
system.

     The problem of binary star clusters can have direct relation to
the galaxies with double nuclei, if assuming that its both components are
supermassive star clusters.

\section{Tidal Radii of Clusters}  %2

     Consider a binary system with globular clusters A and B
of masses $M_{\rm A}$ and $M_{\rm B}$,  mutual distance $a$
and orbital period $P$ by relation
$$
a = 9.49\cdot10^{-8}\,P^{2/3}M^{1/3}\,,                         \eqno(1)
$$
where the summary mass  $M = M_{\rm A} + M_{\rm B}$ is in units of
solar mass, $P$ in days and $a$ in pc. For a binary of globular clusters,
for example, with $M = 10^6\,M_\odot$, and intercomponent distance $a =
300$ pc, from (1) we have for the orbital period: $P = 5\cdot10^8$ years,
i.e. comparable with the dynamical time of the galaxy. During that time
scale such a binary system can perform two-three tens of revolutions
around the mass center of the galaxy.

   However, by reasons which we discuss later, we shall deal with
masses of the order of $10^8 - 10^9\,M_\odot$, i.e. higher as compared with
the mass of ordinary globular clusters in our Galaxy. Adopting therefore,
$M = 10^8\,M_\odot$ and $a = 300$~pc we find from (1): $P = 5\cdot10^7$
years, i.e. shorter as compared with the dynamical time scale of the galaxy.

   The actual sizes of globular clusters in the Galaxy are essentially
determined by the tidal field of the Galaxy. For an isolated globular cluster
the tidal radius $r_{\rm g}$, is determined by the gravitation action of the
Galaxy of a mass $M_{\rm g}$ and radius $R_{\rm g}$; when the cluster is
located in the environments of the Galaxy the tidal radius is given by so
called Hoerner's formula (von Hoerner, 1957):
$$
r_{\rm g} = R_{\rm g}\left(\frac{M_{\rm A}}{2M_{\rm g}}\right)^{1/3}\,.
                                                                \eqno(2)
$$

       As we see, the tidal radius $r_{\rm g}$ of a globular
cluster depends on the masses both of cluster $M_{\rm A}$ and the
Galaxy $M_{\rm g}$. At $M_{\rm A}=10^7\,M_\odot$, for example, for
$M_{\rm g} = 10^{11}\,M_\odot$ and $R_{\rm g} = 15\,000$~pc one has
$r_{\rm g} = 550$~pc.

      The situation is quite different when we deal with a close
binary globular cluster. In this case, the tidal radius
$r_{\rm t}$ for one of the components - the cluster A, is determined
by the gravitation action of the second cluster~B:
$$
\frac{GM_{\rm B}}{(a-r_{\rm t})^2} - \frac{GM_{\rm B}}{a^2} =
\frac{GM_{\rm A}}{r_{\rm t}^2}                              \eqno(3)
$$
or
$$
\frac1{(1-x)^2} - \frac q{x^2} = 1\,,                       \eqno(4)
$$
where $x$ and $q$ denote
$$
x = \frac{r_{\rm t}}a\,.\qquad
q = \frac{M_{\rm A}}{M_{\rm B}}\,.                          \eqno(5)
$$

      Eq.(4) is the correct relationship for the determination of $x$,
i.e. of $r_t$. In particular case, when  $x<<1$, which corresponds
``Galaxy - Globular cluster" combination, we shall have from (4):
$x =(q/2)^{1/3}$ or $r_t = R_g(q/2)^{1/3}$, i.e. the Hoerner's
formula (2).

      As follows from this relationship the tidal radius
$r_{\rm t}$  for the component A depends on the ratio of masses of both
components, but not on the absolute magnitudes of their masses.

     In Table I, the numerical values of tidal radius $r_{\rm t}$, obtained
with the help of Eq.(4), are presented for three values $q= 1, 2, 3$
and the intercomponent distance $a = 500$ pc. As we see, in this case the
tidal radii are 1.5-2 times smaller than those determined by the
gravitational action of the Galaxy. Particularly, for equal masses of both
cluster-components, i.e. when $q = 1$, we have for the tidal radius $r_t =
265$~pc essentially smaller than we have at ``Galaxy - Globular cluster"
combination,i.e. $r_g = 550$~pc.

\begin{table}
{
\renewcommand{\baselinestretch}{1.2}
\renewcommand{\tabcolsep}{10mm}
\begin{center}

                             T a b l e\qquad I
\end{center}
%\vspace{1mm}
{\small
\begin{quote}
        Tidal radius $r_{\rm t}({\rm A})$ for the component A of a binary
        globular
        cluster determined by the gravitation action of the component B
        at three values of mass ratio $q$ and intercomponent distance                                                            ³
        $a = 500$ pc                                                                                                             ³
\end{quote}
\begin{center}
\begin{tabular}{ccc}
\hline
\hline
       $q$   & $x$  & $r_{\rm t}({\rm A})$, pc\\
\hline
        1    & 0.53 &          265            \\
        2    & 0.60 &          300            \\
        3    & 0.65 &          325            \\
\hline
\end{tabular}
\end{center}
}}
\end{table}

     Thus, we arrive at the first remarkable, and important conclusion:
the tidal radius of a globular cluster being a member of a binary cluster
system, is determined by the gravitation attraction of the second
cluster and not of the Galaxy. Such clusters, therefore, are expected
to be more compact than typical isolated globular clusters. In this case
the real sizes of globular clusters are determined not by tidal interaction
of Galaxy mass but by mutual tidal interactions of globular clusters.

\section{Roundchrom of a Binary Globular Cluster?}        %3

     Recently the concept of {\it stellar roundchroms} for close binary
stellar systems has been proposed to explain the observed anomalously
high luminosities of the ultraviolet doublet 2800 MgII of RS CVn type
systems (Gurzadyan, 1996, 1997 ab). Physically the roundchrom is a
common chromosphere enveloping both components of the binary system
having no geometrical contact of their photospheres. The peculiarity of
that concept is in the possibility to provide a large {\it
emission volume} around the secondary component of the binary system.
Geometrical
and physical parameters of such configurations have been revealed for
over 60 RS CVn type close binary stellar systems.

     That idea had inspired us to look for an associated phenomenon for
a binary globular cluster.
As an illustration, in Fig. 1 the configuration of a cluster roundchrom, i.e.
an eight-shaped zero-velocity equipotential
curve enveloping both components of the binary cluster is shown for the
masses of components $M_{\rm A} = 4\cdot10^9\,M_\odot$, $M_{\rm B} =
2\cdot10^9\,M_\odot$, i.e. for the mass ratio $\mu = 0.33$, intercomponent
distance $a = 800$ pc and Jacobi constant $C = 3.94$, so that a narrow
corridor can be formed in the
intermediate Lagrangian point $L_1$ between components. The
dashed circles denote the Roche lobes with radii $R_{\rm L}$ determined
according to the relationship (Eggleton, 1983)
$$
R_{\rm L} = \frac{0.49q^{2/3}}{0.6q^{2/3}+\ln(1+q^{1/3})}\,, \eqno(6)
$$
yielding $R_{\rm L}({\rm A}) = 350$~pc and $R_{\rm L}({\rm B}) = 175$~pc.

      In the case of stars, owing to the existence of strong outer
boundaries -- photospheres, one can compare the star's size with its Roche
lobe. In majority of cases the sizes of main components in close binary
systems coincide with their Roche lobes. Such a coincidence, again as a rule,
is not in the case of secondary components; thus an extensive
emission volume is formed between the outer boundaries of stars and their
Roche lobes.

\vspace{5mm}
\noindent Fig. 1
\vspace{5mm}

In the case of globular clusters the situation is obviously quite different
since one deals with N-body system and hence no clear boundary of the
system exists. Nevertheless, the chaotic properties of the stellar orbits
(see Gurzadyan, Pfenniger, 1994) ensure that basic effects can arise also
for binary globular clusters. Therefore, it appears possible to develop the
principal idea of a roundchrom for the case of binary globular cluster
supposing for simplicity that the linear sizes of the component-clusters are
proportional to $N_{\rm A}$ and $N_{\rm B}$, and the outer boundary of the
first component is coinciding with its Roche lobe, and is somewhat smaller in
the case of the second component.

   Then Figure 1 should be interpreted in the following manner. If the outer
boundary of the component A is not far from the point $L_1$, then one should
expect a transfer or flow of stars from A to B. The surrounding volume, i.e.
the "roundchrom" of the secondary will be occupied mainly by these stars,
resulting a monotonous decrease of the mass of the primary A with corresponding
increase of the mass of the secondary B. As a result, the balance between the
kinetic and potential energies cannot remain frozen for each of the clusters,
along with the secular variation of their linear sizes.

\section{ The Expansion and Contraction of Clusters at Star Transfer Process} %4

     We now aim to determine the rate of the expansion of the
second component-cluster (B) of the binary system, and the rate of
contraction of the first one (A). More definitely, we hope to deduce
the law of the growth of the radius $R_{\rm B}(t)$ of cluster B and the
law of the decreasing of the radius $R_{\rm A}(t)$ by time, i.e. during the
star-transfer process from the first component to the second one.

It is convenient to start from the following basic dynamical principles.
When the moment of inertia of the system can be considered as constant
within some time interval, i.e. its variation is slow enough, one readily
comes to the virial theorem
$$
                    U(t) + 2T(t) = 0\,,                    \eqno(7)
$$
Note, that this relation is obtained not at the infinite time limit, but via
averaging within finite time interval.

Considering the slow stationary star transfer process with conservation
of the moment of inertia (if the
clusters, for example, are far from the last phases of core collapse),
we can assume that the energy conservation condition
$$
                      T_0 + U_0 =  T(t) +  U(t),               \eqno(8)
$$
is fulfilled for each component cluster separately.
Similarly, it is well known that for an isolated cluster
though the evaporated stars carry away some energy, the total energy
of the cluster still can be considered constant within some long enough
time interval while estimating the dynamical characteristics of the cluster.

     Then, we have the well known relationship between radius $R$,
velocity dispersion $v_o$ and the mass $M=mN$ of the cluster
$$
                v_o^2 =\frac G2\,\frac{mN}R\,,                   \eqno(9)
$$
where $N$ is the number of stars, $m$ is the mass of a star.

     Now we can obtain the variations of the sizes for each of these
components of binary cluster A and B during of star transfer process
through the Lagrangian point $L_1$ from the component A to B.

      First consider the behavior of the component B.
For the initial values of kinetic $T_o$ and potential energy $U_o$ we have
$$
T_0 = N_{\rm B}\,\frac{m\,v_0^2}2                       \eqno(10)
$$
$$
U_0 = - \frac G2\,\frac{m^2N_{\rm B}^2}{R_0}\,,           \eqno(11)
$$
where $R_o$ is the initial radius of the cluster B.

    As a result of star transfer process the
cluster B gains an additional kinetic energy
$$
\Delta T = \,{n}t\, \frac{mv_*^2}2\,,                       \eqno(12)
$$
where $v_*$ is the velocity of stars in the transit point $L_1$, and
$n$ is the flux of stars with a constant rate $n$. Thus the total
kinetic energy of cluster B at the moment $t$ will be
$$
T(t)=T_o+\Delta T=N_{\rm B}\,\frac{mv_o^2}2+\,{n}t\,\frac{mv_*^2}2\,.
                                                        \eqno(13)
$$
     Correspondingly we shall have for the potential
energy $U(t)$ of the cluster B at the moment $t$
$$
U(t) = - \frac G2\,\frac{m^2(N_{\rm B}+ nt)^2}{R_t}\,,    \eqno(14)
$$
where $R_t$ is the modified radius of the cluster B.

The condition which we will use is that the virial equilibrium remains valid
during this process, i.e. for moment $t$. This condition has to be safely
fulfilled during the stationary star transfer process due to violent
relaxation effects occuring within dynamical (crossing) time scale.
Then, substituting (13) and (14) into the virial expression (7) and having in
view (9), we obtain for the law of the variation of the radius  $R_t(B)$ of
cluster B by time
$$
\frac{R_t(B)}{R_0}=\frac{(1+\,nt/N_{\rm B})^2}
{1+(v_*/v_0)^2\,{n}t/N_{\rm B}}\,.                       \eqno(15)
$$
This same equation can be obtained using the condition of
conservation of total energy of each of these  clusters during the star
transfer process.

Strictly speaking, $v_*$ must be slightly larger than
$v_0$. However, adopting as a first approximation $v_* = v_0$, we obtain
$$
\frac{R_t(B)}{R_0} = 1+\frac{n}{N_{\rm B}}\,t         \eqno(16)
$$
where the stellar flux rate $n$ is considered constant according to the
assumption on the stationarity of the star transfer process.

     In contrast with cluster B, the cluster A contracts during the stellar
transfer process, i.e. its radius should be decreased. Similar
relationship can be derived for this case as well, i.e. for
the law of decrease of the radius $R_t({\rm A})$ of cluster A
$$
\frac{R_t(A)}{R_0} = 1-\frac n{N_{\rm A}}\,t\,.       \eqno(17)
$$

    However, the star transfer process from the cluster A to the cluster B
cannot proceed stationary too long: with the  decrease of the
diameter of the cluster A, its outer boundary will become more and more
farther from the point $L_1$.  Hence, the initially assumed constant star
transfer rate has to be substituted by a decreasing function
$$
n_t = n_0\,e^{-\gamma t}\,,                                \eqno(18)
$$
where $\gamma$ in $yr^-1$  characterizes the rate of the variations of the
star transfer process from A to B.

      In this case the total number of stars  $n(t)$ transferred from A to B
at the moment $t$, after starting the transfer process $(t = 0)$, will be
$$
n(t) = \frac{n_0}{\gamma}\,(1-e^{-\gamma t})\,.            \eqno(19)
$$

      As a result,
we shall have for the evolution, i.e. for the growth
of the radius of the cluster B at the moment $t$ the following relationship,
again assuming $v_* =v _0$:
$$
\frac{R_B (t)}{R_0} = 1 +\frac{n_0}{\gamma N_{\rm B}}\,(1-e^{-\gamma t})
                                                            \eqno(20)
$$
and for the decrease of the radius of A:
$$
\frac{R_A (t)}{R_0}=1-\frac{n_0}{\gamma N_{\rm A}}\,(1-e^{-\gamma t})\,.
                                                            \eqno(21)
$$

     Above, at the deducation of Eqs. (17) and (21), two extremely case
for the rate of star transfer process are considered, namely, with a
constant rate and with an exponetially decreasing rate. The real situation,
we believe, should be expected between these two limiting cases.

     In Fig.2 we give the curves of time variations for both radii,
$R_{\rm A}(t)$ and $R_{\rm B}(t)$, for the second case, i.e. at exponetially
decreasing star transfer rate given by expression (18), for two sets of initial
parameters, namely, $N_{\rm A} = 5\cdot10^9$, $N_{\rm B} = 2\cdot10^9$,
$\gamma = 2\cdot10^{-9}$~yr$^{-1}$ and $n_0 = 1$, $n_0 = 2$ and
$n_0 = 3$~str.yr$^{-1}$ in the first case (at left), and
$N_{\rm A}=5\cdot10^8$, $N_{\rm B}= 2\cdot10^8$,
$\gamma=10^{-8}$~yr$^{-1}$, and $n_0=0.5$, $n_0=1$ and $n_0=2$~str.yr$^{-1}$
in second (at right).

\vspace{5mm}
\noindent Fig. 2
\vspace{5mm}

       The main difference between both Figures is the time scale,
$10^9$ years in first case and $10^8$ years, in the second. While the
character itself of the evolution of sizes of binary cluster components,
the contraction in one case and expansion in the other, is independent
on rate of the star transfer. Note, that the growth of the radius of the
cluster B takes place faster than the decrease of the radius of A. At the
end of evolution, formally at $t \rightarrow \infty$, the radius of B is
doubled while of A is decreased only on 1/5-th of its initial radius.

     Also, the volume of roundchrom around component B is decreased
mono\-to\-no\-cally during the evolution of the system. The volume of the
roundchrom around A tests an increasing as well however not so strongly.

     It is interesting to note that during the essential evolution of
sizes of cluster A, i.e. during of $2.10^9$ years, the distance between its
outer boundary and the point  $l_1$  is remained practically unchanged. This
important property should be explained also by the slowly motion of the
Lagrangian point  $L_1$ to the direction of the component A synchronously with
the decreasing of its mass.

     Despite of essential differeces in the time scale, in initial total
numbers of stars in clusters as well in the rate of star transfer process,
the main character of curves in Fig.2 is nearly the same. This means that
we cannot say anything for example about the age of binary system or the phase
of evolution or star transfer process of the pair judging only by  direct
image of that or another interacting binary system in the centre of some
galaxies.

    As an illustration, an evolution sequence of the sizes of a binary
cluster is shown in Fig.3. The initial data at $t = 0$  are as follows:
 $N_{\rm A} = 5\cdot10^9$,

\vspace{5mm}
\noindent Fig. 3
\vspace{5mm}

$N_{\rm B} = 2\cdot10^9$ with the same mass (solar) of stars in both
cases, i.e. $m = M_\odot$, the ratio of initial radii of both components
$R_{\rm A}/R_{\rm B} = 2.5$. The configurations of roundchroms
correspond to star transfer rate 2 star.yr$^{-1}$ and $\gamma = 2\cdot10^{-9}$
yr$^{-1}$.
The second configuration from above corresponds to  $t=0.5\cdot10^9$
yr, the third one -- to $t = 2\cdot10^9$ yr.

     In Fig.3, the roundchrom-cluster concentric configurations allow us
 to predict the possible existence of the binary globular clusters,
 particularly in centres of some galaxies, one of components of which in
 two-envelope form, i.e. with an outer low density ring around the
 central dense spheric disk.

     Returning to the star transfer problem, one should outline the
following aspects as well:

    a. One can always find wandering stars in the vicinity of the
corridor in $L_1$, ready for the passage through this corridor.

    b. The gravitation attraction of cluster B, should support the
accumulation of individual stars-members of the cluster A in the vicinity
of the corridor in the point $L_1$,

\section{The Tidal Effect}                       %5

       Independently to the star transfer process from one component to
another, each star, say, in the cluster A, individually undergoes gravitation
attraction by the cluster B. As a result, the star may acquire an additional
velocity.
If so, the macrostructure of energy balance of cluster A must be modified
due to the variation of its linear size.

     To evaluate quantitatively these variations, one has to estimate the
amount of the additional kinetic energy gained due to the gravitation
attraction of the total mass $M_{\rm B}$ of B located in its center
on the distance $a$ from the center of A as is shown in Fig.4.

\vspace{5mm}
\noindent Fig. 4
\vspace{5mm}

    Assuming $N_{\rm A}$ stars of equal mass $m$ homogeneously distributed
within a spherical cluster A with initial radius $R_{\rm A}$ and concentration
$$
n_0=\frac3{4\pi}\,\frac{N_{\rm A}}{R_{\rm A}^3}\,,          \eqno(22)
$$
we can write down the equation of the motion of a star at the distance $r$
from $M_{\rm B}$
$$
m\frac{d\Delta v}{dt}=-G\,\frac{mM_{\rm B}}{r^2}\,,         \eqno(23)
$$
which gives for the additional velocity $\Delta v$ of the star
$$
\Delta v=G\,\frac{M_{\rm B}}{r^2}\,t\,.                     \eqno(24)
$$

     In view of the gravitational action of the cluster B on
individual stars in cluster A, i.e. due to the {\bf tidal} effect, we
can write for the elementary kinetic energy $dT_{\rm tid}$ of a mass
included into the element  of sphere of radius $r$ and thickness $dr$
$$
dT_{\rm tid}=dm\,\frac{\Delta v^2}2=\pi\,n_0\,m\,G^2\,M_{\rm B}^2
(1-\cos\,\varphi)\,\frac{t^2}{r^2}\,dr\,,                   \eqno(25)
$$
where $M_{\rm B} = m\,N_{\rm B}$ and
$$
\cos\,\varphi=\frac{a^2-R_t^2+r^2}{2ar}\,.                  \eqno(26)
$$

     After integration of (25) over limits $a-R_t$ and $a+R_t$, we obtain
for the additional kinetic energy acquired by the cluster A under the
action of cluster B
$$
T_{\rm tid}(t)=\frac34\,G^2m^3\frac{N_{\rm A}N_{\rm B}^2}{R_t^4}\left(
\frac1{u^2-1}-\frac1{2u}\,\ln\,\frac{u+1}{u-1}\right)\,t^2\,,\eqno(27)
$$
where $u=a/R_t$.

    Then the energy conservation law should be written in the form
$$
T_0+U_0=T(t)+U(t)+T_{\rm tid}(t)\,,                          \eqno(28)
$$
where $T_0$ and $U_0$ are the same as above. For $T(t)$ and
$U(t)$ we have
$$
T(t)=\frac G4\,\frac{m^2N_{\rm A}^2}{R_t}\,,            \eqno(29)
$$
$$
U(t)=\frac G2\,\frac{m^2N_{\rm A}^2}{R_t}\,.            \eqno(30)
$$

      Substituting (27), (29) and (30) into (28),and in view of (10) and
(11), we obtain finally the radius $R_t$ of cluster A at the moment of
 time $t$:
$$
\frac{N_{\rm A}}{R_0}\left(1-\frac{R_t}{R_0}\right)=Gm\,\frac{N_{\rm B}^2}
{R_t^4}\left(\frac1{u^2-1}-\frac1{2u}\,\ln\frac{u+1}{u-1}\right)\,t^2
                                                        \eqno(31)
$$
with $u = a/R_t$, and $R_0$ is the initial radius of cluster A.

     The relationship (31) defines the law of the variation of cluster's
radius $R_t$  with time $t$. As we see, the tidal effect of the cluster B
leads to the increase of the radius $R_t$ of cluster A.

     In Fig.5 the curves of the growth of cluster's relative radius,
$R_t/R_0$, with $t$ calculated with the help of (31), are drawn  for three
numerical values of $N_{\rm A}$:
$4\cdot10^8$, $3\cdot10^8$ and $2\cdot10^8$ with $N_{\rm B} = 10^8$ and
$m = M_\odot$, cluster radius $R_0 = 100$ pc and distance between
the centers of clusters $a = 300$ pc.

\vspace{5mm}
\noindent Fig. 5
\vspace{5mm}

     Note, that the growth of the radius $R_t/R_0$ of cluster A due
to the tidal effect  of the cluster B, occurs rather rapidly -- during
a time scale of the order of $10^7$ yrs. This result seems to be in
disagreement with the dynamical nature of the globular clusters.
However, as we will show in the next section, the orbital rotation effect can
essentially suppress the rapid growth of the radius of cluster A.

     The peculiarity of the problem considered above, namely, the role of
the tidal process in binary globular clusters, is that one deals
not with atoms in the stellar atmospheres of interacting close binaries
but with stars.  The main difference is in the absence of collisions
between stars, i.e. the absence of energy exchange process
between stars (particles).

    The next step may be, perhaps, the examination of both processes, the
tidal and star-transfer, simultaneously. However, as shows the preliminary
analysis, the results and basic conclusions will remain unchanged.

\section{The Role of Orbital Rotation}      %6

      Each star in both clusters will undergo the action of
centrifugal force provoked by the orbital motion the components. This
force may compensate in certain degree the gravitational attraction of
the other cluster. This factor, as appears, may change radically
the above conclusions.

      The centrifugal force depends on the tangential velocity $V_t$ and
the distance to the center of rotation of the binary system. Gravitational
attraction provoked by cluster B can be compensated completely by centrifugal
force only if the tangential velocity $V_t$ of a star in the center of
cluster A satisfies the condition
$$
V_t=\left(G\frac{M_{\rm B}}a\right)^{1/2}(1+M_{\rm A}/M_{\rm B})^{-1/2}\,,
                                                        \eqno(32)
$$
where $M_{\rm A}$ and $M_{\rm B}$ are the masses of the clusters,
as before.

     On the other hand, for the orbital velocity $V_{\rm or}$ of a star we
have ( $P$ is the orbital period of the binary system)
$$
V_{\rm or}=\frac{\pi a}P                                \eqno(33)
$$
or using the known expression for $P$,
$$
V_{\rm or}=\frac12\,\left(G\,\frac{M_{\rm B}}a\right)^{1/2}
(1+M_{\rm A}/M_{\rm B})^{-1/2}\,.                       \eqno(34)
$$

      From (34) and (32) we find for the ratio  of both velocities
$$
\frac{V_{\rm or}}{V_t}=\frac12(1+M_{\rm A}/M_{\rm B})\,.\eqno(35)
$$

      In particular case when $M_{\rm A} = M_{\rm B}$, we achieve to
an important result
$$
V_t=V_{\rm or}\,,
$$
i.e. only at equal masses of both clusters we shall have an
equilibrium of both types of forces -- centrifugal and gravitational;
in this case the tidal acceleration of a star towards the cluster A
due to the gravitation attraction of cluster B will be compensated
by the centrifugal force of an orbital rotation of the system. The
further evolution of the sizes of clusters will be determined by star
transfer process from one component to another.

Note, that  depending on the numerical value of the ratio
$M_{\rm A}/M_{\rm B}$ we shall have different relations between $V_t$
and $V_{\rm or}$. So, if $M_{\rm A}/M_{\rm B} > 1$ we have
$$
V_t<V_{\rm or}\,,
$$
and vice versa, when $M_{\rm A}/M_{\rm B} < 1$, we shall have
$$
V_t>V_{\rm or}\,.
$$

      In the first case, when $V_t < V_{\rm or}$, the centrifugal force
will be compensated by the gravitation attraction not completely,
hence, in  this case the tidal acceleration due to the component B may
have some role. Accordingly, some increase of total kinetic energy or
decrease of potential energy in the system A and, correspondingly, a
{\bf decrease} of the linear sizes of cluster  A  may be expected,
although on less degree as compared with the limiting case given by
relationship (21) or curves in Fig.2.

     In the second case when $V_t > V_{\rm or}$, the centrifugal force
will dominate over the gravitational attraction, and the final result
will be the same as in the first case, i.e. we shall have some increase
of kinetic energy or some decrease in linear sizes of the cluster.

    The condition $V_t = V_{\rm or}$ can hardly be fulfilled for
real cluster binaries, therefore the general conclusion may be
formulated in the
following manner: the decrease of linear sizes of the main component of
a binary cluster at star transfer process is inevitable although at a
lower rate as compared with the rate given by (31).

    As to the rate of variation of cluster linear size, it depends in which
degree the additional velocity $V_t$ is larger or smaller than the ``thermal"
(mean) velocity of stars $V_0$ given by relationship
$$
V_0^2=\frac12\,\frac{GM_{\rm A}}{R_{\rm A}}\,.          \eqno(36)
$$

    From (32) and  (36) we find
$$
\frac{V_t}{V_0}=\left(\frac{2R_{\rm A}}a\,\frac{M_{\rm B}}{M_{\rm A}}
\right)^{1/2}\,(1+M_{\rm A}/M_{\rm B})^{-1/2}\,.         \eqno(37)
$$

    For example, at $M_{\rm A}/M_{\rm B} = 2.5$ and $a/R_{\rm A} = 3$ we
have: $V_t/V_0 = 0.28$, i.e. not too small in order to be ignored
and not too large in order to be compared with the results presented in
Fig.4.

    Thus, the star-transfer process from cluster A to cluster B leads to
the {\bf decrease} of the size of cluster A. The tidal effect of the cluster
B on cluster A leads to the {\bf growth} of the sizes of  A.
These are the consequences of the tidal effect in its``pure" form. However,
the rotational effect due to the orbital motion of the clusters A and B,
essentially {\it reduces} the rate of this process. As a result,
the behavior of the size of component A has to be rather complex.

Thus the star-transfer process from cluster A will definitely be accompanied
by the decrease of the cluster's size, however with a  rate essentially
depended on various physical, dynamical and kinematical parameters of both
clusters.

     The latter two effects, the tidal and orbital rotation, lead to changes
in dynamical state of globular cluster during a time scale of the order of
tens of millions years, which is much shorter as compared with the time scale
of globular cluster's evolutionary processes mentioned in Section 1. Thus the
clusters in the binary systems must have different evolutionary time scales
than the isolated globular clusters.

\section{On the Binary Nuclei of Galaxies}          %7

During the last decade the discoveries of galaxies with double nuclei
become more and more common. Due to their form and structure the nuclei
are usually interpreted as binary black holes. Nevertheless the the concept
that nuclei are compact stellar systems cannot be excluded, i.e. a
supermassive globular clusters - of a mass of components of the order of
$10^9\,M_\odot$ and more. An example of such a double nuclei is shown in
Fig.6; this Mrk 273 image is taken by Knapen et all. (1997) via Keck
telescope. This image resembles an eight-shaped binary configuration with
different sizes of components. At the distance of this galaxy 160 million pc,
the intercomponent distance is estimated as 800~pc, with diameters of
components 450~pc and 700~pc.

\vspace{5mm}
\noindent Fig. 6
\vspace{5mm}

    Note that in Fig. 6 the space between the components reveals certain
brightness. According to our concept the point $L_1$ located there
must provide a corridor for star flow from one component to another, so
that the intercomponent space must acquire definite brightness, which can
even be roughly estimated. At the flux {\bf 1 star per year} with stellar
velocity 100 km~s$^{-1}$ we shall have $10^7$ stars during $10^6$ yrs on
10~pc linear length on this area, thus forming a noticeable background
brightness, as maybe indicates Mrk 273.

      At present the number of known galaxies with binary nuclei is over
one hundred. For some of them, including  Mrk 273, the absolute magnitudes
of the components $M_{\rm A}$ and $M_{\rm B}$ have been obtained (Khachikian,
1998). Then, assuming for these nuclei completely stellar composition of
solar mass and solar luminosity, we can evaluate the total number of stars
$N_{\rm A}$ and $N_{\rm B}$ in the components; the results are presented
in Table II.

\begin{table}
{
\renewcommand{\baselinestretch}{1.2}
\renewcommand{\tabcolsep}{5.4mm}
\begin{center}

                             T a b l e\qquad II
\end{center}
%\vspace{1mm}
{\small
\begin{quote}
        Total number of stars $N_{\rm A}$ and $N_{\rm B}$ in the components
        A and B of binary nuclei of six galaxies with known
        absolute magnitudes $M_{\rm A}$ and $M_{\rm B}$ of components
\end{quote}
\begin{center}
\begin{tabular}{cccll}
\hline
\hline
Mrk, No & $M_{\rm A}$& $M_{\rm B}$ & $N_{\rm A}$     & $N_{\rm B}$    \\
\hline
  266   &  - 17.8    &   - 17.5    &\p$2.1\cdot10^9$ &\p$1.6\cdot10^9$\\
  273   &  - 18.4    &   - 17.7    &\p3.6            &\p2.0           \\
  463   &  - 19.5    &   - 19.3    & 10              &\p8.3           \\
  673   &  - 19.6    &   - 19.6    & 10              & 10             \\
  739   &  - 19.1    &   - 18.3    &\p6.9            &\p3.3           \\
  789   &  - 19.5    &   - 17.5    & 10              &\p1.6           \\
\hline
\end{tabular}
\end{center}
}}
\end{table}

      In the case of Mrk 273, the total number of stars in the components
is found $3.6\cdot10^9$ and $2.0\cdot10^9$, i.e. almost of the same order
as those used above when drawing the eight-shape curve in Fig.1. Moreover,
the derived ratio $N_{\rm A}/N_{\rm B} \sim 2$ for this galaxy corresponds,
at least qualitatively, to the photographic images of components in Fig. 6.
Note, the existence of nuclei with equal number of stars in the components
(Mrk 673, ratio 1:1) as well with the ratio up to 1:5 (Mrk 789). It is
remarkable that the total number of stars in the binary nuclei at least
for these six galaxies is of the same order -- $10^9$-$10^{10}$. Even in
view of observational selection, this fact cannot be totally ignored
concerning the nature and composition of the binary nuclei of galaxies.

\vspace{5mm}
\noindent Fig. 7
\vspace{5mm}

      Another example we have in the case of IC 4553 (Arp 220) with a
remarkable binary nuclei structure discovered at radio continuum 4.83
GHz, see Fig. 7 (Baan, Hischick, 1995). The projected separation between
the components is 330 pc, the reconstructed separation 466 pc. At orbital
period of $7\cdot10^6$ yr for the two nuclei with equal masses, the
estimated dynamical mass for both components is $10^{10}\,M_\odot$. It
is important to note, that in this case no clear hot spots are found in
the centers of components within the line emission structure. The spectral
properties confirm the starburst dominated nature of the observed emission,
and hence, the stellar content of both nuclei.

  	The results obtained in present article may find an interesting
application, particularly, for binary globular clusters in the Large
Magellanic Clusters.

\section{Conclusions}                 % 8

      Thus, we formulated and qualitatively considered the {\it
binary system with globular clusters as components} reveals the importance at
least of three  aspects on their dynamics:

    - Mutual tidal interaction of both components.

    - The role of the orbital rotation of the binary system.

    - The star transfer process from one component of the system to the other.

The preliminary conclusions can be summarized as follows:

   $\bullet\,$ Globular cluster as a component of a binary system gains new
dynamical characteristics and properties;

   $\bullet\,$ Dynamical evolution of a globular cluster as a member of
binary system occurs more rapidly as compared with an isolated cluster;

   $\bullet\,$ The tidal radius of a globular cluster in a binary
system is smaller as compared with the tidal radius determined by
the Galactic field;

   $\bullet\,$ The star transfer process from one component to another in
a binary cluster system leads to the reduction of linear size of the first
component and the increase of the size of second one;

   $\bullet\,$ In binary globular clusters the tidal interaction
leads to the increase of
their kinetic energy and, hence, to the expansion of clusters.

   $\bullet\,$ Orbital rotation of a binary cluster system plays a
regularing role on the variations of the sizes of the clusters.

Extensive numerical study of this problem can be of remarkable interest.

I thank V.G.Gurzadyan for valuable discussions.

%\vspace{8mm}
{
\section*{R e f e r e n c e s}

\begin{description}
\item[]\hspace{-1mm} Baan W.A., Haschick A.D. 1995, Ap.J. {\bf454}, 745\\[-6mm]
\item[]\hspace{-1mm} Eggleton P.P. 1983, Ap.J. {\bf268}, 368\\[-6mm]
\item[]\hspace{-1mm} Gurzadyan G.A. 1996a, A\&SS {\bf241}, 211\\[-6mm]
\item[]\hspace{-1mm} Gurzadyan G.A. 1996b, Theory of Interplanetary Flights,
                     Gordon \& Breach,\\[-6mm]
\item[]\hspace{-1mm} Gurzadyan G.A. 1997a, M.N.R.A.S. {\bf290},607\\[-6mm]
\item[]\hspace{-1mm} Gurzadyan G.A. 1997b, New Astr. {\bf2}, 31\\[-6mm]
\item[]\hspace{-1mm} Gurzadyan V.G., Pfenniger D. (Eds.) Ergodic Properties in
                     Stellar Dynamics, Springer-Verlag, 1994.
\item[]\hspace{-1mm} Hoerner S. von 1957, Ap.J. {\bf125}, 451\\[-6mm]
\item[]\hspace{-1mm} Khachikian E.Ye.1998, Trans.Armenian Nat.Acad.Sci. {\bf
                     98}, 239\\[-6mm]
\item[]\hspace{-1mm} Knapen J.H., Laine S., Yates J.A., Robinson A., et al.,
                     1997, Ap.J.Lett. {\bf 490}, L29\\[-6mm]
\end{description}
}

\section*{Figure Captions}

{\small
\begin{description}
\item[Fig.1.] Roundchrom structure for a close binary system with star
         clusters as components with mass ratio $\mu = 0.33$, masses of
         components $M_{\rm A}=4\cdot10^9 M_\odot$, $M_{\rm B} = 2\cdot10^9
         M_\odot$ and intercomponent distance $a=800$ pc, $R_{\rm L}$(A)
         and $R_{\rm L}$(B) are the radii of Roche lobes. All linear sizes
         are in parsecs.

\item[Fig.2] The time-dependent curves of the variations of linear radii
          $R_t / R_0$ of both components, A and B, of binary system with
          globular clusters as components. For the initial moment, $t=0$, it
          is assumed $R_t/R_0=1$. The curves are calculated for three values of
          star-transfer rate  $n_o$ from $0.5$ up to 3 str.yr$^{-1}$ and for two
          values of  gamma, $2.10^{-9}$ yr$^{-1}$ and $10^{-8}$ yr$^{-1}$
          and the time scale  $10^9$ yr and $10^8$ yr. During the
          star-transfer process from component A to component B,
          component A undergoes a decrease of its radius, i.e. $R_t/R_0 < 1$
          (lower half) and the component B, the contrary, an increase,
          $R_t / R_0 > 1$ (upper half).

\item[Fig.3] Three stages of the evolution of globular clusters' diameter
         as a function of time $t$, i.e. during star-transfer process from
         component A to component B through the corridor in Lagrangian
         point $L_1$.  All three configurations are calculated for one and the
         same star-flow rate $n_o = 2$ str.yr$^{-1}$ and
         $gamma = 2.10^{-9}$ yr$^{-1}$.
         Initially, at $t = 0$, the component B has a powerfull roundchrom
         (pointed area) and component A is practically without a roundchrom.

\item[Fig.4] To the calculations of the tidal effect of
         the cluster B of a mass $M_{\rm B}$ on the stars in cluster A.

\item[Fig.5] The estimated time-depended variations of radius,
         $R_t/R_0$, of component A due to the tidal effect for
         three values of total number of stars $N_{\rm A}$ in
         the cluster A, at its radius $R_0 = 100$ pc and intercomponent
         distance $a = 300$ pc.

\item[Fig.6] Double nuclei of Mrk 273 (Knapen et al 1997). Components of
         nuclei are of different masses and different sizes.
         The separation between  components is 730~pc. Note the
          background in the corridor between components.

\item[Fig.7] Double nuclei of IC 4553 in 4.83 GHz
         (Baan and Haschick 1995). The separation between
         nuclei is 330 pc.
\end{description}
}

\end{document}